\documentclass[twocolumn,a4paper]{article}
\usepackage{graphicx}
\usepackage{algorithm}
\usepackage{algorithmic}

\begin{document}

\title{Reliable Data Storage in Distributed Hash Tables}

\author{Matthew Leslie, Oxford University}

\date{June 2005}

\maketitle

\begin{abstract}

Distributed Hash Tables offer a resilient lookup service for
unstable distributed environments. Resilient data storage,
however, requires additional data replication and maintenance
algorithms. These algorithms can have an impact on both the
performance and the scalability of the system. In this paper, we
describe the goals and design space of these replication
algorithms.

We examine an existing replication algorithm and present a new
analysis of its reliability. We then present a new dynamic
replication algorithm which can operate in unstable environments. We
give several possible replica placement strategies for this
algorithm, and show how they impact reliability and performance.

Finally we compare all replication algorithms through simulation,
showing quantitatively the difference between their bandwidth use,
fault tolerance and performance.

\end{abstract}

\section{Introduction}

Grid Computing and Content Distribution applications place
demanding scalability and fault tolerance requirements on
underlying services. Client-Server based solutions have often
encountered problems meeting these requirements
\cite{scalability1,scalability2}.

Peer to peer technology has demonstrated excellent scalability in
file sharing applications\cite{measurep2p}. It seems likely this
potential will be invaluable in solving scalability problems for
more legitimate grid service applications.

Much recent research in the area of peer to peer computing has
focused around various kinds of distributed hash table (DHT).
These algorithms provide scalable fault tolerant key based
routing. Key lookups can be routed reliably to the host
responsible for them, which return any associated data.

In this paper we explore the potential of Distributed Hash Tables
to provide reliable, scalable and consistent storage of mutable
data. We show how the choice of replication algorithm can affect
the performance, reliability and bandwidth costs of storing
data in a Hash Table, a topic that has previously received little
attention.

We will first give a brief overview of distributed hash tables,
and describe the aims and design space of replication algorithms.

We then describe both an existing replication algorithm and a new
replication algorithm based on dynamic replication\cite{dynamic},
but adapted to provide reliability in an unstable environment,
investigating the performance and reliability of both.

We then provide an analysis of the impact of various replica
placement strategies possible with our dynamic replication
algorithm. We show that replica placement can have a significant
impact on the reliability and performance of the system.

Finally, we use a simulation to give a quantitative comparison of
the performance and bandwidth costs of the algorithms we have
described.

\section{Distributed Hash Tables}

Distributed hash tables provide a solution to the lookup problem
in distributed systems. Given the name of a data item stored in
the system, we can locate the node on which that data item should
be stored. Most DHTs aim to find the responsible node with delay
logarithmic in the number of nodes in the network.

There are many different DHT implementations, including
PAST\cite{past}, Tapestry\cite{tapestry}, CAN\cite{CAN},
Kademlia\cite{kademlia} and Chord/DHash \cite{chord}. A node in a
DHT is typically responsible for the data close to it according to
some distance function. Each node maintains knowledge about a
small proportion of other nodes in the network, and uses these to
forward requests for keys it does not own to nodes which are
closer to the requested key. Usually, the distance is reduced by a
constant fraction at each routing hop, leading to lookup time
logarithmic in the number of nodes.

In any large scale distributed system, nodes are likely to be
joining and leaving the system constantly. These changes in the
set of participating nodes are called \emph{churn}. DHT routing is
often tolerant of node churn, but storage is not. When a node
fails, the key-value pairs it was responsible for become
inaccessible, and must be recovered from elsewhere. This means
that to provide reliability a \emph{replication algorithm} must
store and maintain backup copies of data. It must do this in a
manner scalable both in the number of nodes and the quantity of
data stored in the DHT.

In this paper, we will discuss replication in the context of
Chord. Briefly, Chord nodes and data items are given IDs between 0
and some maximum $K$, which map to positions on a ring.  The
distance function between IDs is the clockwise distance around the
ring between them.  A node \emph{owns}, or is responsible for,
data that it is the first clockwise successor of. In order to
route key requests, each node maintains knowledge of its immediate
clockwise neighbors, called its \emph{successors}, and several
other nodes at fractional distances around the ring from it,
called its \emph{fingers}. Space concerns prohibit giving a full
description of Chord, and for a full description readers are
encouraged to consult the original paper\cite{chord}

\section{The Aims of Replication}
\label{aims}
 A Replication algorithm aims to achieve some
combination of the following design goals.

\begin{description}
\item[Reliability] The replication algorithm must not rely on any
single node, and must recover from churn without user
intervention.

\item[Scalability] The replication algorithm should scale to
storing large quantities of data on a large number of nodes, $N$.
So as not to limit the scalability of Chord, per node replication
algorithm state and bandwidth usage should be $O(log(N))$.

\item[Lookup latency] The replication algorithm may reduce the
time taken to look up information by placing replicas of data in a
manner that allows network locality to be exploited.

\item[Mutability] Updating data involves enumerating all replicas.
Once this is done, a  distributed commit protocol can be used to
update the data consistently at all locations. For this reason,
enumerating all replicas should be as fast as possible.

\item[Load balancing] The replication algorithm may provide caches
of more popular items in order that the load is evenly balanced
among all the nodes in the network.

\item[Consistency] If the replication algorithm is to work with
mutable data, it should seek to provide clear guarantees about the
consistency of replicas.

\end{description}

Different algorithms achieve these aims to varying extents. The
choice of replication strategy may depend on which goals are more
important to the task being considered.

Other aims may include resilience to malicious nodes, anonymity or
privacy. These are important goals, but we consider them
orthogonal to the replication problem, and best treated
separately.

\section{Replication Algorithms}

A replication algorithms can be characterized by its approach to
four key problems:

\begin{description}

\item[Replica Maintenance] Node churn will cause replicas to be
lost. The replication algorithm must detect and repair these lost
replicas without using excessive bandwidth.

\item[Replica Addressability] In order to update an item, we need
to locate all replicas of that item. Ways of doing this include
limiting replica placement to a fixed number of nodes, searching
for replicas, and periodically deleting old replicas.

\item[Replica Placement]  The replica placement strategy
determines which nodes replicas should be stored on. This can have
an impact on both performance and reliability.

\item[Replica Cardinality]The number of replicas we keep of a
given key may either be fixed in advance, or allowed to vary
according to the key's popularity. Variable cardinality often
provides superior load balancing, but makes addressability more
difficult to achieve.

\end{description}

In this paper, we consider replication by storing a complete copy
of the data associated with each key on another node. We believe
that the algorithms we will describe could, with some adaptation,
also be applied to erasure coded fragments of the original data,
with possible performance benefits in some
circumstances\cite{erasureVsReplication}.

 We will now give details of how two replication algorithms
approach these tasks.

%
%
%
%
%

\section{DHash Replication}
\label{succ-rep}

This replication algorithm is a combination of the replica placement
strategy first proposed in the original Chord paper\cite{chord}, and
a maintenance algorithm proposed in \cite{chord:cates-meng}. The
techniques are combined in the DHash\cite{dhash} storage system.

The placement strategy is simple, replicas of a data item are
placed on the $r$ successors of the node responsible for that data
item and nowhere else.

Newly joined nodes will inherit a keyspace from the node they
precede, and are sent the data they become responsible for when the
maintenance algorithm next runs.

The DHash maintenance algorithm runs two maintenance protocols in
order to prevent the number of replicas from either dropping too
low or rising too high:

\begin{description}
\item[Local Maintenance] A node sends a SYNCHRONIZE message
containing the key range it is responsible for to its $r$
successors. These nodes then synchronize the contents of their
databases so that all keys in this range are stored on both the
root node and its successors.

Efficient methods for database synchronization, such as Merkle
Tree hashing\cite{Merkle1980}, are discussed in
\cite{chord:cates-meng}.

\item[Global Maintenance] A node periodically checks its database
of keys to see if it has any keys it is no longer responsible for.
To do this, it looks up the owner of each key it stores, and
checks the successor list of that owner.

If it is within $r$ hops of the node, then it will be within the
first $r$ items of the successor list.

If its ID is not in this list, the node is no longer responsible
for keeping a replica of this item, and the node offers the data
item to the current owner. It can then safely delete the key.

\end{description}

If all replicas are to be repaired by a single maintenance call,
the local maintenance algorithm must run two passes, the
first gathering all replicas in the root nodes key range onto the
root node, the second distributing these replicas onto all
successors.

\subsection{Fetch Algorithm}

If we adopt this replication algorithm, we can use the following
fetch algorithm to retrieve the data associated with a key. This algorithm
will also share the load of providing a data  item between all the replica holders.

\begin{algorithm}
\caption{Fetch for $key$ under Successor Replication}
\label{dhashfetch}
\begin{algorithmic}
\STATE $successors \leftarrow findSuccessors(key)$

\WHILE{$\neg item$ \textbf{and} $\neg successors.isEmpty()$}
    \STATE $node \leftarrow successors.popRandom()$
    \STATE $item \leftarrow node.get(key)$
\ENDWHILE

\end{algorithmic}
\end{algorithm}

\subsection{Maintenance and Reliability}
\label{select-rep-int}

In order to keep the system reliable, we must both store replicas
and repair them regularly. The more replicas we keep, the less
frequently maintenance is required. This means that, for a given
level of reliability, there is a trade-off between the bandwidth
used by maintenance algorithm runs and the disk space used for
storing replicas.

Here, we attempt to give some new insight into this trade-off, and
show the level of replication and maintenance necessary to provide a
given level of reliability.

For a given data item to be lost, all $r$ of the nodes holding
replicas of it must fail. If a replica is missing from the system
with probability $p$, the probability of a given data item being
lost is simply $p^r$.

For the purposes of providing reliable storage, however, we are
concerned not with the probability of a given item being lost, but
with the probability of data loss anywhere in the system. For this
to happen, a node and its $r$ successors must all have failed at
some point in the ring.

To determine this probability, we model the chord ring as a sequence
of $N$ nodes, each of which is missing replicas with probability
$p$. The probability of data loss in this model is equivalent to
that of obtaining a sequence of $r$ successful outcomes in $N$
Bernoulli trials with probability of success $p$. This is known as
the \emph{Run Problem}, and the general solution, $RUN(p,r,N)$, can
be given in terms of a generating function\cite{runMathworld}.

\begin{eqnarray*}
 F(p,r,s) & = & \frac{p^r s^r (1-ps)}{1-s+(1-p)p^rs^{r+1}} \equiv
 \sum_{i=r}^{\infty}c^{p}_{i}s^{i} \\
 RUN(p,r,N)  & = & \sum_{i=r}^{N}c_{i}^{p}
\end{eqnarray*}

In order to relate this to maintenance intervals, we need to
determine the proportion of replicas missing from the ring. We
will do this in terms of the number of maintenance calls in one
\emph{half life}, $S$. A network's half life is the minimum of the
times taken for $N/2$ of the nodes to leave, and the time $N/2$
nodes take to join. We will consider only stable state systems,
where nodes join and leave the network at the same rate.

After one half-life, half of the nodes are newly joined, and
contain no data\footnote{This is also a simplification. Although
most missing data is caused by empty new nodes, a node also stores
data on its $r$ successors, and so a failure causes one fewer
replica to be stored. Taking this into account gives the fraction
of missing replicas as $\frac{2r+1}{4rS}$. Our approximation is
fair for large $r$.}, so we take $p = \frac{1}{2}$. We assume
churn occurs at a constant rate, so that a fraction $\frac{1}{S}$
of the way through a half life, $p=\frac{1}{2S}$.

We make the simplifying assumption that data transfer time is a
negligible proportion of the half life, so that a maintenance call
will instantly return the system to its ideal state, provided no
data has been lost completely before it runs. Each half life can
then be divided into $S$ independent maintenance intervals, during
each of which data is lost with probability $RUN(\frac{1}{2S},r,N)$.
The overall data loss probability is the probability that any of
these $S$ maintenance intervals loses data, and so is given by:

\begin{eqnarray*}
FAIL(N,r,S)=1 - ( 1 - RUN(\frac{1}{2S},r,N))^S
\end{eqnarray*}

\begin{figure}
 \centering
 \includegraphics{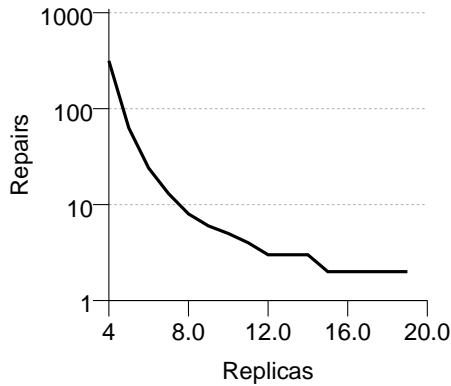}
 \caption{Minimum repairs necessary to maintain $FAIL(N,r,S)=10^{-6}$ for $N=500$}
 \label{repairs_vs_replicas}
\end{figure}

We can use this to determine how often the maintenance algorithm
needs to run in order to maintain a given failure probability.
Figure \ref{repairs_vs_replicas} shows the minimum maintenance
frequency necessary to maintain $FAIL(500,r,S)=10^{-6}$, where $r$
varies from 4 to 20. We can see that there is a clear trade-off
between bandwidth use and storage space. The number of maintenance
calls necessary drops rapidly as we increase the number of
replicas.

\begin{figure}
 \centering
 \includegraphics{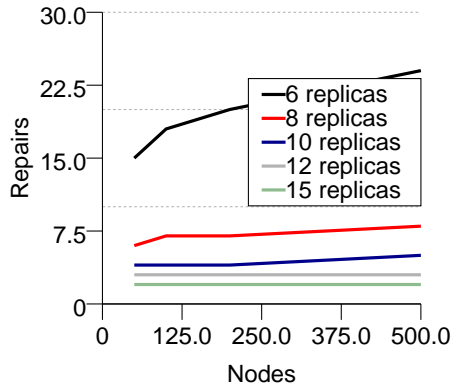}
 \caption{Minimum repairs necessary to maintain $FAIL(N,r,S)=10^{-6}$ in networks
of size $N={50,100,200,500}$, with $r={6,8,10,12,15}$}
 \label{repairs_vs_nodes}
\end{figure}

Figure \ref{repairs_vs_nodes} shows how network size affects the
level of maintenance necessary. $N$ is varied from 50 to 500 for
several values of $r$. For small $r$, the network size has a
significant impact on the required maintenance frequency. As we
increase $r$, network size becomes less important in determining
maintenance frequency.

It should be noted that in systems where data size is large or
half-life is short, so that data transfer times become a
significant fraction of a half life, our assumptions are not valid
and more frequent maintenance or a larger numbers of replicas will
be necessary.

\subsection{Maintenance and Fetch Latency}

\label{dynreppeformance} As we allow the number of replicas to
drop between maintenance intervals, we increase the likelihood
that we will need to contact more than one node to find the data.
If we again approximate the probability of a node not having data
as $\frac{1}{2S}$ then the number of nodes we need to contact, or
probe, before data is found will be geometrically distributed,
with expected value:

\begin{eqnarray*}
 E(probes) = \frac{2S}{2S-1}
\end{eqnarray*}

\section{Dynamic Replication}
\label{dynrep} Replica Enumeration, as proposed in \cite{dynamic},
aims to remove some of the placement and cardinality restrictions
imposed by successor replication, whilst preserving addressability
and the ability to make consistent updates.

The placement strategy for Replica enumeration is based around an
\emph{allocation function}, $h(m,d)$. For each document with ID $d$,
the replicas are placed at replica locations determined by $h(m,d)$
where $m\geq1$ is the index of that instance. The allocation
function is intended to be pseudo-random, so that the replicas are
evenly distributed about the address space.

The replication cardinality is variable in a fixed range $1 \leq r
\leq R_{MAX}$, allowing greater replication for items in greater
demand. The mechanism used to decide the exact number of replicas
is not specified, but could be designed to adapt to both the
reliability of the network, and the load on those nodes providing
each document.

To provide addressability, the following invariants are
maintained:

\begin{enumerate}
    \item Replicas of an item $d$ are only placed
at addresses given by $h(m,d),$ where $1 \leq m \leq R_{MAX}$.
    \item For any document $d$ in the system, there always exists
an initial replica at $h(1,d)$
    \item Any further replica with $(m>1)$ can only exist if a
replica currently exists for $m-1$
\end{enumerate}

Various strategies for finding data are possible in this scheme.
One that is generally efficient is to do a binary search over the
range [$1..R_{MAX}$] starting from a randomly selected index in
that range. If the data is not replicated at a given location, we
use invariant three to reduce the search range accordingly.

Dynamic replication can help alleviate the lookup bottleneck
 that affects Successor replication. Successor replication
requires that all lookups for a popular key are directed to that
key's owner. With an appropriate allocation function, dynamic
replication can place replicas at evenly spread, well known,
locations around the ring.  Lookup queries for the owner of a
popular item are then distributed more evenly.

\subsection{Maintenance algorithm}
A major difficulty with the dynamic replication algorithm as
originally provided is that maintaining these invariants in a
system with a high churn rate is very difficult. Node departures
and arrivals could cause any of the invariants to be violated.

It can be shown that lookups will proceed correctly as long as at
least invariant two holds. However, for replica addressability,
invariants one and four must also hold.



We modify the algorithm given in \cite{dynamic} with a maintenance
algorithm to allow it to operate correctly and reliably in a
system with high churn rates.

We will use the following definitions to refer to the various
roles nodes play in holding replicas:
\begin{enumerate}
\item  The node responsible for $h(1,d)$ is the \emph{owner} of
$d$. \item  The \emph{replica group} for an item $d$ are those
nodes whose keyspace includes a replica location from the set $ \{
h(m,d) : 1\leq m \leq R_{MAX}\}$.

\item  The \emph{core group} for an item $d$ is the set of replica
holders for which $ m \leq R_{MIN}$

\item  The \emph{peripheral group} are those replica holders for
which $R_{MIN} < m  \leq R_{MAX}$.
\end{enumerate}

Since we can not rely on any single node being
available in an unreliable environment, we must modify our invariants.

\begin{enumerate}
    \item Replicas of an item $d$ can only be retrieved from addresses given by $h(m,d)$ where
    $\{1\leq m \leq R_{MAX}\}$.
    \item For any document $d$ in the system, there exists with high probability
a replica in the core group.
    \item Any peripheral replica with $(m>R_{MIN})$ can only be retrieved for a single maintenance interval if no
replica currently exists for $m-1$.
\end{enumerate}

We can now give three maintenance protocols which maintain these
invariants under churn.

\begin{description}
\item[Core Maintenance] The owner of a data item calculates and
looks up the nodes in the core group for the data range it is
responsible for. For each core replica holder, the owner and the
replica holder synchronize databases over the part of owner's
keyspace which is mapped into that replica holder's keyspace.

Core maintenance must also deal with \emph{allocation collisions},
as described in the next section.

\item[Peripheral Maintenance]

In order to maintain Invariant 3, a node which stores a replica
with index $m>R_{MIN}$ must check that a replica of that item is
also held on the replica predecessor, the owner of the location
with index $m-1$.

For each peripheral replica a node holds, it must obtain a summary
of the items with the previous index on the replica predecessor.
Bloom filters\cite{bloom} can be used to reduce the cost of these
exchanges.

These summaries can be used to remove orphaned peripheral replicas
from the system. Orphaned peripheral replicas should not be used
to answer fetch requests, but should still be stored for at least
one maintenance interval, as simulation shows maintenance often
replaces the missing replica.

\item[Global Maintenance] Each node calculates the replica group
for each item it holds. Any items it is no longer a replica holder
for are offered to their owner, then deleted.
\end{description}

We cache lookups made during maintenance to reduce bandwidth
costs. Cache validity is checked at regular intervals during
maintenance.

This algorithm attempts to restore the system to its ideal state
each time it is run. However, between runs, the system is rarely in
its ideal state. Thus we must ensure the system operates correctly
where the invariants do not hold. Invariant two is sufficient to
ensure lookups proceed correctly, though less efficiently, as nodes
fail\cite{dynamic}.

In order to update information, we must discover how many
peripheral replicas are currently in the system. To be completely
certain of consistency, we must offer all updates to all owners of
peripheral replica locations.

If a lower probability of a temporary inconsistency is acceptable,
we can improve performance by offering updates only until we
encounter a certain number of empty peripheral replica locations
since, by invariant three, it becomes increasingly unlikely that any
further locations are occupied. This could dramatically improve
performance if $R_{MAX} \gg R_{MIN}$.

\subsection{Allocation Collisions}
\label{collisions}

It is important that the images of a node's key-range under
$h(m,d)$ are owned by different nodes for each $d$. In some cases
however, the allocation function will map two replicas into the
keyspace of the same node. We call this an \emph{allocation
collision}. Each Allocation collision reduces the number of nodes
in the core replica group, reducing reliability.

The core maintenance algorithm must keep track of which nodes have
been allocated replicas from which key ranges. If an allocation
collision occurs, the core maintenance algorithm must instead
place the collided keyspace must in the peripheral group,
effectively extending the core group. This means we must choose
$R_{MAX}$ so that sufficient peripheral locations are available to
recover from all allocation collisions with high probability. To
do this, we must understand how nodes are distributed around the
ring.

Since we have $N$ nodes uniformly distributed throughout a
keyspace of size $K$, the probability of a node being at a given
ID is $\frac{K}{N}$. Therefore, the number of keys between nodes
is geometrically distributed with $p=\frac{K}{N}$. Using standard
properties of a geometric distibution\cite{geometric}, we can find
the mean and variance of this distribution.

\begin{eqnarray*}
\mu = \frac{K-N}{N} \cong \frac{K}{N} \: (since\: K \gg N)\\
\sigma^2 =\frac{K(K-1)}{N^2} \cong \frac{K^2}{N^2} \: (since\: K
\gg 1)\
\end{eqnarray*}

To recover from allocation collisions, the range of available
replica locations should be at least this size. Thus, by the central
limit theorem\cite{limit}, the space occupied by $r$ nodes will be
normally distributed with

\begin{eqnarray*}
\mu \cong \frac{rK}{N} \\
\sigma^2 \cong \frac{rK^2}{N^2}
\end{eqnarray*}

And so, by standard properties of the normal distribution, 95\% of
the time, the keyspace between $r$ nodes will be less than
$(r+1.645\sqrt{r})\frac{K}{N}$ keys in length. To allow $r$ replicas
can be stored in 95\% of cases, the range of available replica
locations should be at least this size.

\subsection{Dynamic fetch algorithm}

The dynamic fetch algorithm needs to choose which indexes to
lookup and in which order. In many cases, algorithm \ref{fetch}
will give good performance.

\begin{algorithm}
\caption{Dynamic Fetch for $key$} \label{fetch}
\begin{algorithmic}
\STATE $indexes \leftarrow [0 \ldots R_{MAX}]$

\STATE $item\leftarrow NULL$

 \WHILE{$\neg item$}
    \STATE $index \leftarrow indexes.popRandom()$
    \STATE $item \leftarrow recursiveGet(key,h(index,key))$
    \IF {$\neg item$ \textbf{and} $index > R_{MIN}$}
        \STATE $indexes.removeRange(index,R_{MAX})$
    \ENDIF
    \IF {indexes=[]}
        \STATE $indexes \leftarrow [0 \ldots R_{MAX}]$
    \ENDIF
\ENDWHILE

\end{algorithmic}
\end{algorithm}

In situations where load balancing is not critical, shorter fetch
times can be attained by searching the core replica locations
before trying peripheral ones.

If maintenance is infrequent, eliminating the entire range of
peripheral replicas with higher indexes if one peripheral replica
is found empty may result in poor performance, and simply removing
the replica known to be empty may be preferable.

\subsection{Recursive data lookup}

In order to increase performance when looking up data, we use
algorithm \ref{recget} to perform recursive gets. This combines the
Chord lookup and get messages, which allows any node on the lookup
path of a request to return a replica of the requested data, if it
holds one. This avoids further lookup hops, reducing fetch latency.

\begin{algorithm}

\caption{Recursive get for $key$ at $location = h(m,key)$ for some
m} \label{recget}
\begin{algorithmic}

\IF{$self.keyrange.containsReplica(key)$ \textbf{and}
    $self.store.contains(key)$  }
    \STATE \textbf{return} $(self.store.get(key))$
\ENDIF

\IF{self.keyrange.contains(location)}
    \STATE \textbf{return} $NULL$
\ENDIF

\STATE $next =
self.closestPrecedingNode\footnotemark[2]{}(location)$

\STATE \textbf{return} $next.recursiveGet(key,location)$
\end{algorithmic}
\end{algorithm}

Recursive data lookup can interfere with load balancing, since
some replicas are passed queries more often than others. To
prevent this, an overloaded node may choose to forward a recursive
get request rather than answer using its own replica.

\subsection{Allocation Functions}

 \label{allocfunc}

The choice of allocation function is critical to maintenance
performance. For each item $d$ that a node owns, it must lookup
and contact every node in the core replica group in order to run
the core maintenance protocol.

\footnotetext[2]{ closestPrecedingNode as described in
\cite{chord}}

In order that this is scalable, we must ensure that as many of
these lookups as possible can be satisfied with little network
communication. This requires that the allocation function maps one
nodes data onto a limited number of replica holders.

We suggest that for a given $m$, $h(m,d)$ is a \emph{translation}
in d. This means that the image of one node's key-range is another
continuous linear range of the same size. Since Chord nodes are
equally distributed throughout the key space, an image of one
node's key-range is owned by $O(1)$ other nodes.

We will now give four allocation functions and explore how they
impact reliability and performance. All these functions make use
of $N$, the number of nodes in the system. This value may either
be supplied by the user, or estimated at run time\cite{chordsize}.

\subsubsection{Successor Allocation}
\label{dynsucc} $h(m,d) = (d + (m \cdot \frac{K}{N})) \:
\textbf{mod} \: K$\\

Attempting to map replicas onto successors is efficient, as the
Chord Protocol maintains a list of each nodes successors on that
node, so lookups can often be performed without consulting another
node.

Because replica locations with different indices are relatively
close together under this allocation function, we expect some
allocation collisions, which must result in the creation of
peripheral replicas. Using the result from section
\ref{collisions}, we recommend that $R_{MAX} - R_{MIN}$ is at
least $1.645\sqrt{R_{MIN}}$, so that $R_{MIN}$ distinct replicas
can be stored in 95\% of cases.  This consideration also applied
to the predecessor and block allocation functions.

\subsubsection{Predecessor Allocation}
 \label{dynpred}

$h(m,d) = (d-(m \cdot \frac{K}{N}))\: \textbf{mod}
\: K$\\

Because queries are routed around the ring clockwise towards the
node responsible for them, a lookup for one node is frequently
routed through one of its predecessors.

Predecessor allocation aims to exploit this fact to reduce lookup
latency. When a request for a document is routed through one of
the replica holders, the recursive get algorithm allows them to
satisfy the request before it ever reaches the node responsible,
reducing the fetch latency by one or more hops.

\subsubsection{Block Allocation}
\begin{eqnarray*}
h(m,d) = (d - (d \: \textbf{mod} \: \frac{K\cdot R_{MAX}}{N}) \\+
(d \: \textbf{mod} \: \frac{K}{N}) \\ + (m*\frac{K}{N})) \:
\textbf{mod} \: K
\end{eqnarray*}

This allocation function attempts to make replica groups overlap
entirely with core replica groups of other nearby keyranges. As
will be seen in the next section, this policy provides a lower
probability of data loss than other placement functions.

It also provides most of the benefits of both successor and
predecessor replication, since most nodes will have replicas
placed on both successors and predecessors.

This allocation function is discontinuous in $d$, and the
maintenance algorithm must be able to deal with this when mapping
ranges of keys onto other nodes.

\subsubsection{Finger Allocation}
\label{dynfing}

\begin{eqnarray*}
\delta = log_2(\frac{K}{N})\\
h(m,d) = (d+2^{(m+\delta)}) \: \textbf{mod} \: K\\
\end{eqnarray*}

This allocation function again takes advantage of the information
already maintained by the chord algorithm. Chord maintains routing
information about nodes at fractional distances around the ring,
called fingers. Placing replicas on these finger nodes reduces the
number of lookups that must be made, and distributes replicas
evenly around the ring.

Because of the distance between replica locations, allocation
collisions are rare, and $R_{MAX}$ may be set more conservatively
than for other allocation functions.

\subsection{Allocation functions and Reliability}

The allocation function chosen has a significant impact on
reliability. Block allocation, in which core replica groups for
one data range overlap entirely with core groups for nearby data
ranges, produces only a very few core replica groups. Finger
allocation produces core replica groups which overlap very little
with other replica groups, producing a large number of distinct
groups.

\begin{figure}
 \centering
 \includegraphics{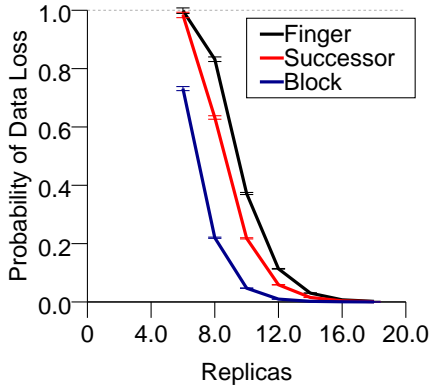}
 \caption{Probability of failure in one halflife for three allocation functions under 50\% failure.  Predecessor allocation is not shown as its performance is equivalent to successor. Error bars show 95\% confidence intervals. }
 \label{reliability}
\end{figure}

This large number of distinct groups leads to a higher probability
that any one of them will fail. We produced a simple model of a
500 node network, in which 250 nodes are marked as failed. We
produced $10^5$ sample networks with this model, and used them to
estimate the probability of any data loss occurring in the network
with varying numbers of replicas. Figure \ref{reliability} shows
the probability of data loss for finger, block and successor
allocation functions. Block allocation is able to achieve a more
reliable system with a smaller number of replicas.

We also simulated random replica placement, where replicas were
placed entirely randomly using a pseudo random number generator.
The results for random placement are almost identical to those for
finger placement.

\begin{figure}
 \centering
 \includegraphics{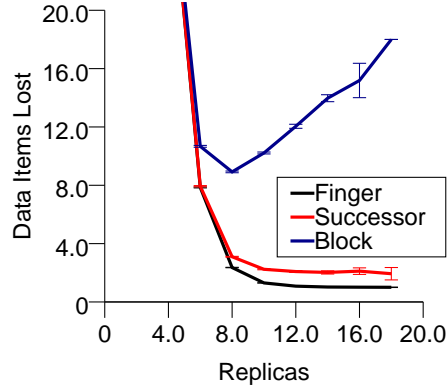}
 \caption{Quantity of data lost in failures for Successor, Random and Finger allocation functions. Error bars show 95\% confidence levels. }
 \label{dataloss}
\end{figure}

In figure \ref{dataloss}, we compare the quantity of data lost,
given that a failure occurs. More data is lost with block
allocation functions when a rare failure does occur than other
allocation functions. In many applications however, any quantity
of data loss would be considered catastrophic.

\section{Simulation}

We now attempt order to quantitatively compare the performance and
bandwidth usage of these replication algorithms. Due to the
difficulty of managing large numbers of physical nodes\cite{SWORD},
we chose to test the algorithms through simulation rather than
through deployment.

Our simulator is based around the SimPy\cite{simpy} discrete-event
simulation framework, which uses generator functions rather than
full threads to achieve scalability. The simulator implements a
message level model of a Chord network running each of the
replication algorithms described.

\begin{figure}
 \centering
 \includegraphics{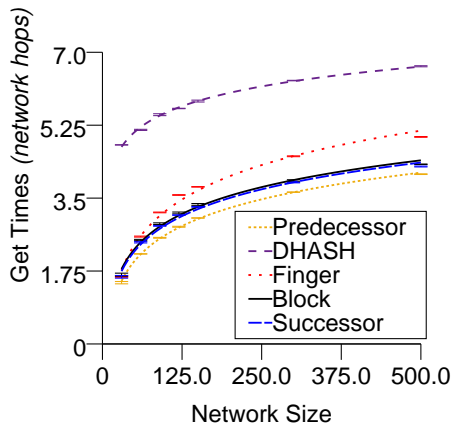}
 \caption{Get times in various system sizes. (Successor and Block overlap).}
 \label{latency_vs_size}
\end{figure}

\subsection{Simulation Parameters}
In our simulation, we chose parameters that might resemble a data
center built from cheap commodity components. While the simulator
is capable of running thousands of nodes, we have limited it to
two hundred in most scenarios. This was in order to keep runtime
reasonable for the large number of scenarios and algorithms we
wish to test, and the need to repeat simulations to estimate
errors.

We simulate a steady state system in which a node fails every 24
hours, shortly after which a new node replaces it. Latency between
nodes is assumed to be uniform, and bandwidth is assumed to be
unlimited - messages always take the same time to deliver
regardless of size.

We chose fixed parameters for the chord algorithm in all
simulations, with a successor list length of 10 and finger table
size of 12. Chord repair is carried out at thirty minute
intervals.


We also configure the GET algorithm to search the core replica
group before trying the peripheral replicas. Local and Core
Maintenance algorithms run two passes at each maintenance
interval.

Failures are detected by timeouts, which are set to 3 hops for
round trip communications. Recursive lookup timeouts are based on
network size, and are set to 15 hops in a 200 node network. A
shorter timeout could have been chosen, leading to shorter average
lookup times, as failed lookups are detected more quickly. Short
timeouts also increase bandwidth usage, however, as long-running
lookups are reissued before they complete.

The system is simulated for one complete half life, during which
50,000 sample data fetches are made for randomly chosen data
items, and fetch times are logged. Bandwidth usage is also logged
by type, allowing separation of maintenance messages from chord
repair messages. Each simulation is repeated 4 times to obtain a
good estimate of the average latencies and bandwidth usage.

\section{Simulation Results}

\subsection{Fetch Latency}
The fetch latency each maintenance algorithm can achieve depends
on the network size, the frequency with which it is run, and the
number of replicas in the system.
\begin{figure}
 \centering
 \includegraphics{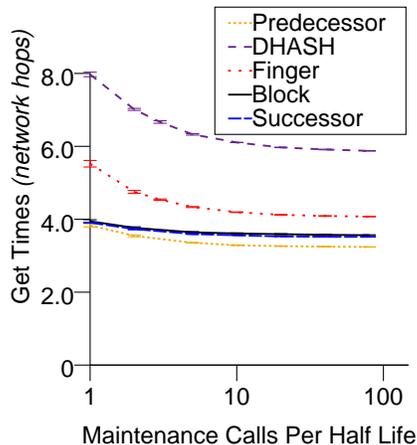}
 \caption{Fetch times in a 200 node system, varying $S$, the number of times the maintenance algorithm runs per half life. (Successor and Block overlap).}
 \label{latencyVsS}
\end{figure}

Figure \ref{latencyVsS} shows how fetch times scale with
maintenance frequency. The Finger and DHash algorithms both scale
as predicted in section \ref{dynreppeformance}.

Maintenance frequency has less effect on Successor, Predecessor and
Block fetch times. The proximity of different replica indexes means
that other replica holders often preemptively return data, so that
it is less important that specified replica is present.

The predecessor algorithm achieves the shortest fetch times. With
predecessor allocation, queries for core replicas are more often
routed through peripheral replicas, which return the data
preemptively. DHash fetches are never returned preemptively, and
are routed through the requesting node, so that they take at least
one hop longer than dynamic allocation.

Figure \ref{latency_vs_size} shows how fetch times scale with
network size. All algorithms show logarithmic behavior, though
finger allocation compares increasingly badly to other dynamic
algorithms since as networks size grows, the probability of a
preemptive return drops.

Increasing the number of replicas reduces the lookup times
slightly, whereas increasing the number of distinct data items in
the system has no impact on lookup times.

\subsection{Maintenance Bandwidth Costs}
\begin{figure}[h]
 \centering
 \includegraphics{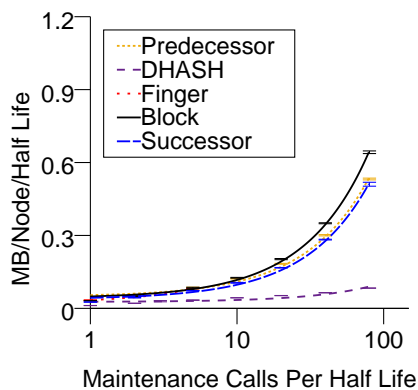}
 \caption{Overhead bandwidth in a 200 Node system with varying numbers of repairs}
 \label{S_vs_oh}
\end{figure}
\begin{figure}[h]
 \centering
 \includegraphics{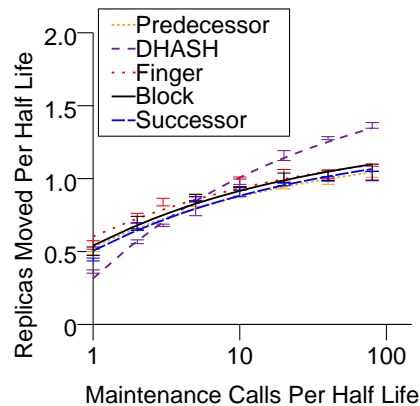}
 \caption{Proportion of data in system transmitted in one halflife. (Predecessor, Block and Successor overlap) }
 \label{S_vs_movement}
\end{figure}

Maintenance bandwidth can be divided into two separate costs: The
cost of identifying which data should be stored where, and the
cost of moving the data to that location. We refer to the former
as the maintenance \emph{overhead}.

In figure \ref{S_vs_oh} we can see how overhead varies with
maintenance frequency. DHash maintenance has lower overhead than the
dynamic algorithms. Dynamic maintenance typically involves $O(r)$
lookups, where DHash requires $O(1)$. The dynamic algorithms all
have similar overhead, which increases linearly with maintenance
frequency. Notably, all algorithms overhead bandwidth is so small as
to be negligible in most network environments.

Data movement bandwidth is likely to be the bottleneck in
distributed storage systems. Figure \ref{S_vs_movement} shows all
dynamic algorithms move very similar quantities of data. At high
maintenance levels, significantly more data is moved with DHash than
the dynamic algorithms, since a single node joining produces changes
in the membership of $r$ nearby replica groups, with one node
leaving each group. This causes the node expelled from each of these
replica groups to send any replicas it no longer owns to their
owner.


\begin{figure}
 \centering
 \includegraphics{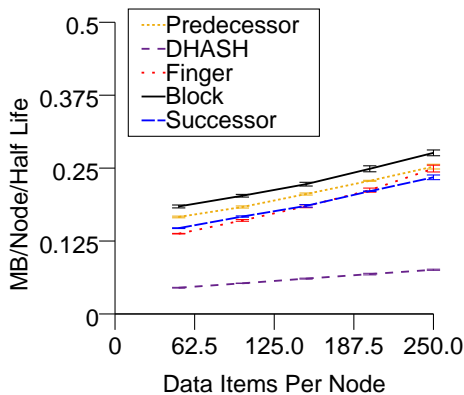}
 \caption{Overhead bandwidth in a 200 Node system with varying numbers of data items per node}
 \label{items_vs_oh}
\end{figure}

Figure \ref{items_vs_oh} shows the linear relationship between the
number of key value pairs and maintenance overhead. Again,
maintenance overhead is low for all algorithms, so that it should
be feasible to store a very large numbers of items per node
without maintenance bandwidth becoming a bottleneck.

Per node maintenance overhead bandwidth and total data movement
remain constant as we vary the number of nodes in the system, for
all algorithms. This means all algorithms will scale well to very
large numbers of nodes.

\subsection{Fault Tolerance}
We have so far investigated the performance of these data
replication algorithms under a steady state of churn, in which new
nodes join at the same rate as other nodes fail. The DHT can also
recover from far higher failure rates, although there is a
substantial performance impact.

\begin{figure}
 \centering
 \includegraphics{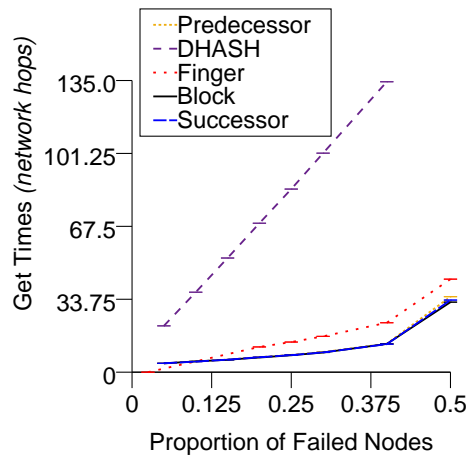}
 \caption{Data fetch times after a simultaneous failure of varying proportions of nodes. (Block Successor and Predecessor overlap)}
 \label{catastrophe}
\end{figure}

We simulate a simultaneous failure of a varying proportion of the
nodes in the chord network, and then launch 50,000 data fetches
immediately afterward. Unlimited retries are allowed, and the
average fetch time, including retries is shown in Figure
\ref{catastrophe}.

The DHash algorithm is particularly affected in this scenario, due
to its reliance on reaching a single node. The dynamic algorithms
are more resilient to faulty routing, as they may select multiple
indexes to look up, and because preemptive returns are possible
even when the specific node requested is unreachable.

\section{Related Work}

Metadata based algorithms, such the Version ID system used by
OceanStore\cite{oceanstore} are another possible replica management
option. These algorithms do not have placement restrictions, but
instead use a metadata item to locate replicas. These solutions may
perform well in many scenarios, but require an underlying reliable
storage system for metadata. The algorithms we have described could
be used to provide such a storage system.

Soft-state storage is another common replica management system. With
soft-state storage, no attempt is made to maintain replicas.
Instead, data expires after some timeout and the system relies on
data periodically being refreshed and reinserted by some external
system. Though this can be useful for frequently refreshed data,
failure of the external storage system can cause unwanted data loss.

\section{Conclusions}
We have used a combination of analysis and simulation to assess
the ability of various replication algorithms to meet the goals we
set out in section \ref{aims}.

We can see that dynamic replication can achieve faster lookups,
greater reliability and may require less replica movement than the
DHash algorithm, with only a slightly higher maintenance overhead.
We have also shown how the allocation function choice can have a
dramatic impact on performance. Of the allocation functions we
considered, block allocation provides the best reliability and
represents a good compromise for most systems, though predecessor
placement might be preferable if performance is critical.

Possible drawbacks of dynamic replication are its slightly higher
maintenance bandwidth usage and its reliance on an even distribution
of node IDs, which may make it unsuitable for small Chord Rings.

System size scalability is good for all maintenance algorithms.
Lookup times scale with the logarithm of network size and total
system bandwidth consumption scales linearly with the number of
nodes.

On an internet wide scale, bandwidth and uptime are likely to be
more limited than in the data center scenarios we have considered.
In such a system, the number of nodes is likely to vary throughout
the day rather than remain constant. Although our work may provide
insight into performance in such scenarios, further work needs to be
done to assess the reliability of a system which incorporates user
desktop systems.

\nocite{past,cfs} \small
\bibliography{replication}
\bibliographystyle{plain}
\end{document}